\documentclass[pra,aps,twocolumn,superscriptaddress,notitlepage]{revtex4-2}
\usepackage{amssymb,amsfonts,amsmath,amstext,graphicx,hyperref,lipsum,empheq}
\usepackage{float}
\usepackage[normalem]{ulem}
\hypersetup{
	colorlinks=true,
	linkcolor=blue,
	filecolor=magenta,      
	urlcolor=cyan,
}
\usepackage{amsthm}
\usepackage{outlines}

\usepackage{outlines}
\usepackage{subfigure}
\usepackage{cleveref}

\usepackage[dvipsnames]{xcolor}

\newcommand{\eref}[1]{(\ref{#1})}

\newcommand{\beq}[0]{\begin{equation}}
\newcommand{\eeq}[0]{\end{equation}}

\def\be{\begin{equation}}
\def\ee{\end{equation}}
\def\bea{\begin{eqnarray}}
\def\eea{\end{eqnarray}}
\newcommand{\ba}{\begin{eqnarray}}
\newcommand{\ea}{\end{eqnarray}}
\usepackage{hyperref}

\def\ket#1{|{#1}\rangle}
\def\braket#1{\langle{#1}\rangle}
\def\Bra#1{\left\langle#1\right|}
\def\Ket#1{\left|#1\right \rangle}
\def\BraVert{\egroup\,\mid\,\bgroup}

\definecolor{myblue}{rgb}{.8, .8, 1}

\begin{document}

\title{Collective effects enhanced multi-qubit information engines}
\author{Noufal Jaseem}%
\affiliation{Department of Physical Sciences, Indian Institutes of Science Education and Research Berhampur, India.}

\author{Victor Mukherjee}%
\affiliation{Department of Physical Sciences, Indian Institutes of Science Education and Research Berhampur, India.}

\date{\today}

\begin{abstract}
 We study a quantum information engine (QIE) modeled by a multi-qubit working medium (WM) collectively coupled to a single thermal bath.  We show that one can harness the collective effects to significantly enhance the performance of the QIE, as compared to equivalent engines lacking collective effects.  We use one bit of information about the WM magnetization to extract work from the thermal bath.   We analyze the work output, noise-to-signal ratio and thermodynamic uncertainty relation (TUR), and contrast these performance metrics of a collective QIE with that of an engine whose WM qubits are coupled independently to a thermal bath. We show that in the limit of high temperatures of the thermal bath,  a collective QIE always outperforms its independent counterpart. In contrast to quantum heat engines, where collective enhancement in specific heat plays a direct role in improving the performance of the engines, here 
the collective advantage stems from higher occupation probabilities for the higher energy levels of the positive magnetization states, as compared to the independent case. Furthermore, we show that in the limit of high temperatures, the TUR for both the collective as well as the independent QIEs violate the standard heat engine TUR bound of $2$.
\end{abstract}

\maketitle

%%%%%%%%%%%%%%%%%%%%%%%%%%%%%%%%%%%%%%%%%%%%%%%%%%%%%%%
\section{Introduction}
The relationship between information theory and thermodynamics has been a topic of intense interest in the research community during the last few decades \cite{vinjanampathy2016quantum}, with information engines, i.e. engines which use information to extract useful work, playing a significant role in this regard \cite{kim2011quantum, diaz14quantum, aydin20landauer}.  The concept of Maxwell's demon~\cite{knott1911life}, first proposed by James Clerk Maxwell in 1867, was one of the earliest works suggesting the possibility of using information to extract work from a single heat bath in classical systems. 
Subsequently, the idea of using information and measurements to extract work was extended to the quantum regime \cite{kim2011quantum,aydin20landauer,lutz2015maxwell,jussiau2022many,rio2011thermodynamic,myers2022quantum,vinjanampathy2016quantum,yanik2022thermodynamics,elouard2017extracting, bhandari2022continuous,kammerlander2016coherence,jordan2020quantum,bresque2021two,manikandan2022efficiently,buffoni2019quantum}, and also realized experimentally in recent years \cite{koski2014experimental, kunkun22experimental}.

  One of the major aims of the field of quantum thermodynamics  is to study the effects of quantum physics on the operation of quantum machines  \cite{klimovsky15thermodynamics, bhattacharjee21quantum, mukherjee21many,cangemi2023quantum}. For example, quantum coherence has been shown to improve the power output of quantum heat engines \cite{uzdin15equivalence}, while squeezed thermal reservoirs have been used to significantly enhance the efficiencies of such engines \cite{rossnagel14nanoscale, niedenzu18quantum, jan17squeezed}; non-adiabatic excitations in quantum critical systems driven out of equilibrium have been shown to aid in the charging of quantum batteries \cite{obinna22harnessing}, and entanglement may help in extracting work from such batteries \cite{alicki13entanglement}. Quantum features have been used to design quantum information engines (QIEs) as well. For example, a bosonic quantum Szilard engine has been shown to outperform its fermionic counterpart \cite{kim2011quantum}. Ref. \cite{kammerlander2016coherence} showcases the use of quantum coherence to extract non-trivial work from the process of unselective measurements, akin to information erasure. This stands in contrast to the Landauer principle, which imposes a thermodynamic work cost for both classical and quantum erasure.
%In Ref. \cite{kammerlander2016coherence}, quantum coherence was used to enhance the output work of a QIE.

In recent years, collective effects in many-body quantum technologies comprising indistinguishable particles collectively coupled to thermal baths have received significant attention from the research community \cite{niedenzu18cooperative, kloc19collective,  latune20collective, souza2021collective, jaseem2022quadratic, dmytro23performance}. For example, collective effects have been used to perform high-precision quantum thermometry~\cite{latune20collective} and to obtain enhanced work output~\cite{niedenzu18cooperative, kloc19collective,  latune20collective}, efficiency~\cite{klimovsky19cooperative}, and reliability~\cite{jaseem2022quadratic} in quantum thermal machines. In this work, we propose a collective effects enhanced many-body QIE, modeled by a working medium (WM) of $n$ non-interacting indistinguishable spin $1/2$s, collectively coupled to a single heat bath. Information about the magnetization direction of the WM allows us to extract work. Our analysis shows that in comparison to an equivalent QIE modeled by  $n$ non-interacting spin $1/2$s independently coupled to heat baths, the collective effects may result in significantly enhanced work output, low noise-to -signal ratio, and low Thermodynamic Uncertainty ($\mathcal{Q}$). The collective advantage originates from higher occupation probabilities for the higher energy levels of the positive magnetization states compared to the independent case. 

The paper is organized as follows. In Sec. \ref{sec:engine}, we discuss the model of the many-body QIE, its operation, and the process of work extraction, in detail. The  different performance metrics of the QIE are studied, and compared with their independent QIE counterparts, in Sec. \ref{sec:work_statistics}. Finally, we summarize our results in Sec.~\ref{sec_discussion}.

%%%%%%%%%%%%%%%%%%%%%%%%%%%%%%%%%%%%%%%%%%%%%%%%%%%%%%%
\section{Many-body information engine}
\label{sec:engine}

\subsection{Information engine}
\label{sec:infoengine}

\begin{figure}[!htb] 
    \centering
    \includegraphics[width=0.85\linewidth]{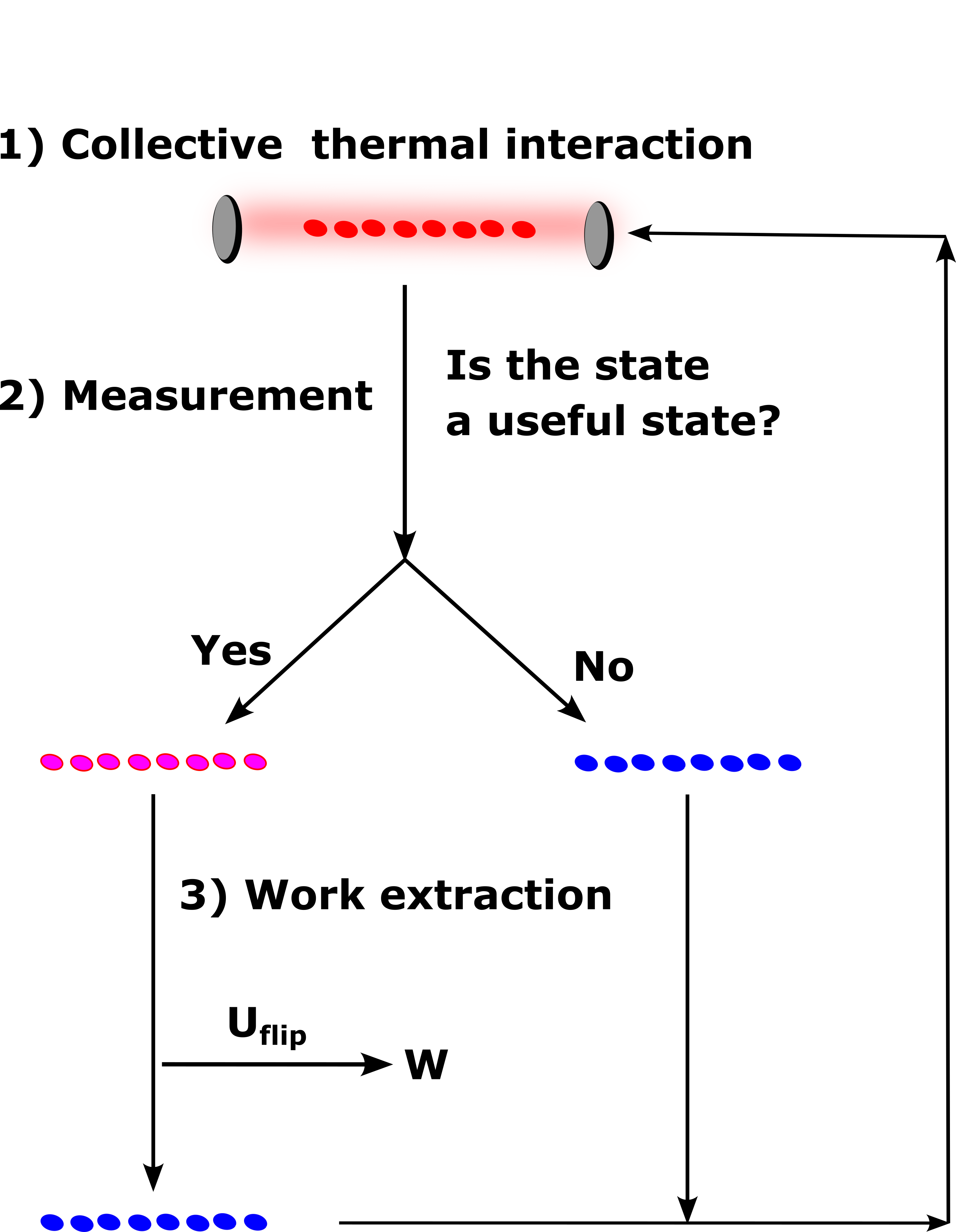}
    \caption{ A schematic representation of an information engine comprising $n$ qubits, highlighting the three strokes of the cycle.}
    \label{fig:info_engine}
\end{figure}

We consider a working medium (WM) consisting of $n$ spin-$1/2$ systems. One can model a three-stroke information engine as follows (see Fig. \ref{fig:info_engine}):
\begin{itemize}
\item Stroke 1: In the first energizing stroke, the WM is coupled to a thermal bath at an inverse temperature $\beta$.  The WM undergoes a non-unitary evolution, wherein energy flows from the bath to the system, resulting in the WM reaching a steady state $\rho^{\text{ss}}$ at long times. 

\item Stroke 2: In the second stroke  the WM is decoupled from the bath, and the direction of the magnetization $m$ of the WM, or equivalently, the sign of $m$, is measured. A positive magnetization implies the WM has non-zero ergotropy, i.e., the maximum work that can be extracted  via suitable  unitary operations \cite{campaioli18quantum}. A part of the ergotropy can be extracted by applying a global unitary $U_{\rm flip}$ to flip all the qubits of the WM.

\item Stroke 3: Following the result of the measurement in stroke 2, we apply a global unitary $U_{\rm flip}$   to flip all the qubits of the WM for $m > 0$, thereby changing the sign of $m$, and extracting work in the process.  On the other hand, a negative  $m$ is detrimental to work extraction under the above unitary $U_{\rm flip}$. Consequently  no unitary is applied in case of $m \leq 0$.
\end{itemize}
Finally, we couple the WM with the Markovian bath introduced in stroke 1, and repeat all the steps mentioned above, thereby giving rise to a cyclic  information engine. 

\subsection{Model and dynamics}
\label{sec:model}

In this section, we discuss in detail the dynamics involved in stroke 1 of the information engine, introduced above. To this end, we consider a WM  composed of multiple non-interacting  qubits with bare Hamiltonian given by $H=\hbar \omega \mathcal{J}_z$, where we define the collective angular momentum operators as $\mathcal{J}_i := \frac{1}{2}\sum_{k=1}^n \sigma^k_{i}$. Here  the Pauli matrix $\sigma_i^k$  is the spin operator along the $i$ axis ($i = x, y,z$), acting on  the $k$-th spin. The spins are collectively coupled to a thermal bath during the non-unitary stroke 1 through the system-bath interaction Hamiltonian $H_{\rm int} = \gamma \mathcal{J}_x \otimes \mathcal{B}$. Here $\gamma$ denotes the coupling strength, while $\mathcal{B}$ is an operator acting on the thermal bath. As shown in the references~\cite{niedenzu18cooperative, latune2019thermodynamics, latune20collective}, such collective system-bath interaction may generate quantum coherence in the steady-state density matrix $\rho^{\rm ss}_{\rm col}$ of the WM, characterized by the presence of off-diagonal elements in the WM density matrix in the local spin basis. The dynamics of the WM is described by the master equation~\cite{niedenzu18cooperative, latune2019thermodynamics, latune20collective}
\begin{eqnarray}
\frac{d\rho}{dt}=-\frac{i}{\hbar}[H,\rho]+\Gamma(\omega)\mathcal{D}(\mathcal{J}^+)\rho+\Gamma(-\omega) \mathcal{D}(\mathcal{J}^-)\rho,
\label{eqmaster}
\end{eqnarray}
where $\mathcal{D}[O]\rho =( 2 O^\dagger \rho O- O O^\dagger \rho - \rho O O^\dagger)/2$, the collective ladder operators of the spin system are given by $\mathcal{\mathcal{J}}^\pm := \mathcal{\mathcal{J}}_x \pm i \mathcal{\mathcal{J}}_y$, and $\Gamma(\omega)$ is the spectral function of the bath at frequency $\omega$. The spectral functions $\Gamma(\omega)$ and $\Gamma(-\omega)$ are related to each other via the Kubo-Martin-Schwinger condition $\Gamma(-\omega) = \exp\left(-\beta\hbar \omega \right)\Gamma(\omega)$ \cite{breuer02}. The steady-state $\rho^{ss}_{\rm col}$ can be expressed in the collective basis $|j,m\rangle_i$ of the common eigenvectors of $\mathcal{J}_z$ and $\mathcal{J}^2=\mathcal{J}_x^2+\mathcal{J}_y^2+\mathcal{J}_z^2$  as~\cite{latune2019thermodynamics}:
\begin{eqnarray}\label{eq:rhoss}
\rho^{ss}_{\rm col}(\beta,\omega)&=&\sum_{j=j_0}^{n/2}\sum_{i=1}^{l_j}P_{j,i}\; \rho_{j,i}^\text{th}(\beta,\omega),\\
\rho_{j,i}^\text{th}(\beta,\omega) &=& \frac{1}{Z_j} \sum_{m=-j}^j e^{-\beta m\hbar\omega} |j,m\rangle_i { }_i\langle j,m|.\nonumber
\end{eqnarray}
Here $Z_j=\sum_{m=-j}^j e^{-\beta m\hbar\omega}$ is the partition function of the angular momentum subspace $j$, $-j\leq m\leq j$, $j\in [j_0; n/2]$, and $j_0=1/2$ for odd $n$ and $j_0 = 0$ for even $n$. The index $i\in [1; l_j]$, where $l_j$ is the multiplicity of the eigenspaces associated with the eigenvalue $j$ of the operator $\mathcal{J}^2$~\cite{mandel1995optical}. The initial state, $\rho_0$, of the WM determines the probability $P_{j,i}$ through the relation $P_{j,i}=\sum_{m=-j}^j\; { }_{i}\langle j,m|\rho_0|j,m\rangle_i$. 

Previous studies have shown that an initial state prepared in the  $j = n/2$ subspace is the most favorable for obtaining collective advantage in quantum thermal machines \cite{niedenzu18cooperative, latune20collective, jaseem2022quadratic}. Consequently, here we restrict ourselves to the $j = n/2$ subspace, such that $l_{j = n/2} = 1$, $P_{j=n/2}=1$ and the corresponding steady state is given by \cite{latune20collective, mukherjee21many}
\begin{equation}\label{eq:dicke_ss}
     \rho^{ss}_{\rm col}(\beta,\omega)= \frac{1}{Z_{n/2}} \sum_{m=-n/2}^{n/2} {e^{-\beta m\hbar\omega}} \Ket{\frac{n}{2},m}\Bra{\frac{n}{2},m}.
\end{equation}

 We note that in contrast to the setup introduced above, one can consider an independent WM-bath coupling, described by an interaction Hamiltonian of the form $H_{\rm int} = \frac{\gamma}{2} \sum_{k=1}^n \sigma_x^k \otimes \mathcal{B}_k$, where $\mathcal{B}_k$ denotes an operator acting on the bath coupled to the $k$-th spin.  In the latter scenario, the steady-state reached at the end of the non-unitary thermal interaction stroke is a direct product state of the form ~\cite{breuer02, niedenzu18cooperative, latune2019thermodynamics, latune20collective, mukherjee21many}
\begin{eqnarray}
\rho^{ss}_{\rm ind}(\beta,\omega) &=&
\otimes_{k=1}^n \rho^{ss}_{k}(\beta,\omega),
\label{eq:thermalstate}
\end{eqnarray}
where $\rho^{ss}_{k}(\beta,\omega) = \frac{\exp \left(-\beta \hbar \omega \sigma_z^k/2\right)}{{\rm Tr}\left[\exp \left(-\beta \hbar \omega \sigma_z^k/2\right)\right]}$ is the steady state density matrix corresponding to the $k$th spin.

%%%%%%%%%%%%%%%%%%%%%%%%%%%%%%%%%%%%%%%%%%%%%%%%%%%%%%%
\section{Statistics of the output work} \label{sec:work_statistics}
The unitary operation $U_{\rm flip}$ considered in the second stroke introduced in Sec.~\ref{sec:infoengine} acts only the states $|n/2,m\rangle$ with $m>0$; a part of the ergotropy of such states is extracted in the form of output work, while those states are transformed to $|n/2, -m\rangle$ at the end of this stroke.
For the independent bath coupling 
as well, a state $\ket{m}$ with $m > 0$ denotes the presence of non-zero ergotropy, which can be used to perform useful work through unitary transformation to the state $\ket{-m}$.  
We take the sign of energy outflow from the WM as positive and the inflow as negative. If $w_m=2 m\hbar\omega$ is the amount of work extracted from the state $\ket{n/2,m}$ or from $\ket{m}$ with probability $p_m^{\alpha}$ ($\alpha =$ col., ind.), then the average work takes the form,
\begin{eqnarray}\label{eq:w}
\langle W_{\alpha} \rangle &=& \sum_{m\geq 0} w_m p_m^{\alpha},
\end{eqnarray}
and its variance, $\text{var}(W_{\alpha})=\braket{W_{\alpha}^2}-\braket{W_{\alpha}}^2$, is given by
\begin{eqnarray}\label{eq:varw}
\text{var}(W_{\alpha}) &=& \sum_{m\geq 0} w_m^2 p_m^{\alpha} - \langle W_{\alpha} \rangle^2.
\end{eqnarray}
In Eqs. \eqref{eq:w} and \eqref{eq:varw}, $m=0,1,2,...,n/2$ for even $n$, while  $m=1/2,3/2,5/2,...,n/2$ for odd $n$.  
For the collective engine, the probability is given by (see Eq. \eqref{eq:dicke_ss})
\begin{eqnarray}
p_m^{\rm col}= \frac{e^{-\beta m\hbar\omega}}{Z_{n/2}},
\label{eq:pmcol}
\end{eqnarray}
 while in the case of the independent engine, we have (see Eq. \eqref{eq:thermalstate})
 \begin{eqnarray} 
 p_m^{\rm ind}= \;{}^nC_{n/2-m}\;\frac{e^{-\beta m\hbar\omega}}{Z_{\rm ind}}; \qquad  ^rC_s\equiv\frac{r!}{(r-s)! s!}.
 \label{eq:pmind}
 \end{eqnarray}
 Here $Z_{\rm ind}=\left(e^{-\beta  \omega /2}+e^{\beta  \omega /2}\right)^n$ is the partition function corresponding to the thermal state of the independent engine's WM. The factor $^nC_{n/2-m}$ in Eq.~\eqref{eq:pmind} denotes the number of combinations of spin arrangements that give rise to a magnetization $m$.

 \begin{figure}[!htb]
    \centering
    \includegraphics[width=0.97\linewidth]{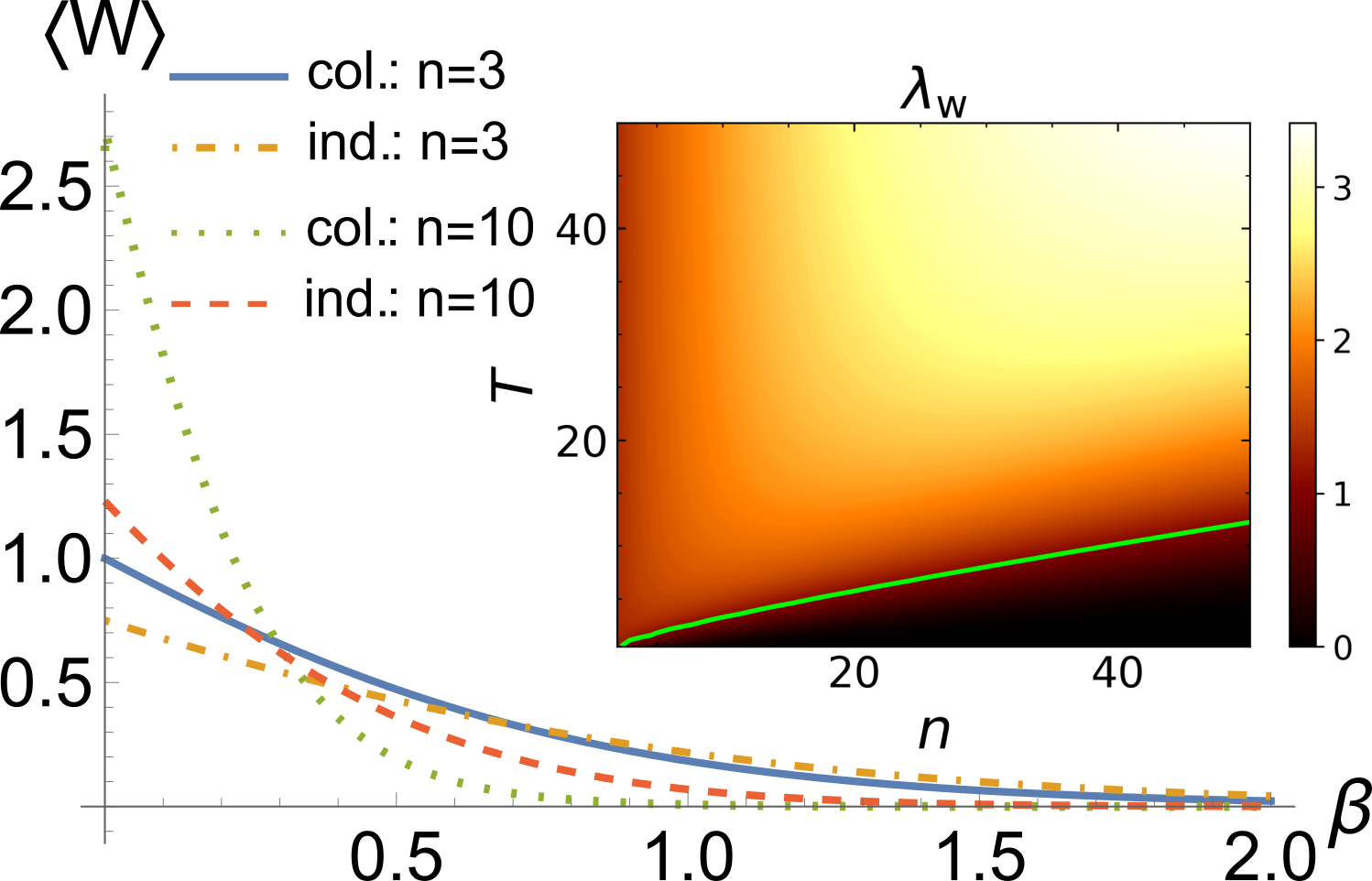} %wvsT.png
    \caption{The average work outputs $\langle W_{\rm col}\rangle$ and $\langle W_{\rm ind}\rangle$ are plotted as functions of $\beta$ for $n=3$ and $n=10$. In the inset, we present the variation of the ratio $\lambda_{\rm w}=\langle W_{\rm col}\rangle / \langle W_{\rm ind}\rangle$ with respect to $n$ and $T=1/\beta$. The green contour line represents $\lambda_{\rm w}=1$. The values of $\hbar$, $k_{\rm B}$ and $\omega$ are set to 1.}
    \label{fig:wvsT}
    \end{figure}

        %%%%%%%%%%%%%%%%%%%%%%%%%%%%%%%%%%%%%%%%%%%%%%%%%%%%%%%%%%%%%%%%%%%%%%%%%%%%%
 \begin{figure}[t]
     \centering
     \includegraphics[width=0.48\textwidth]{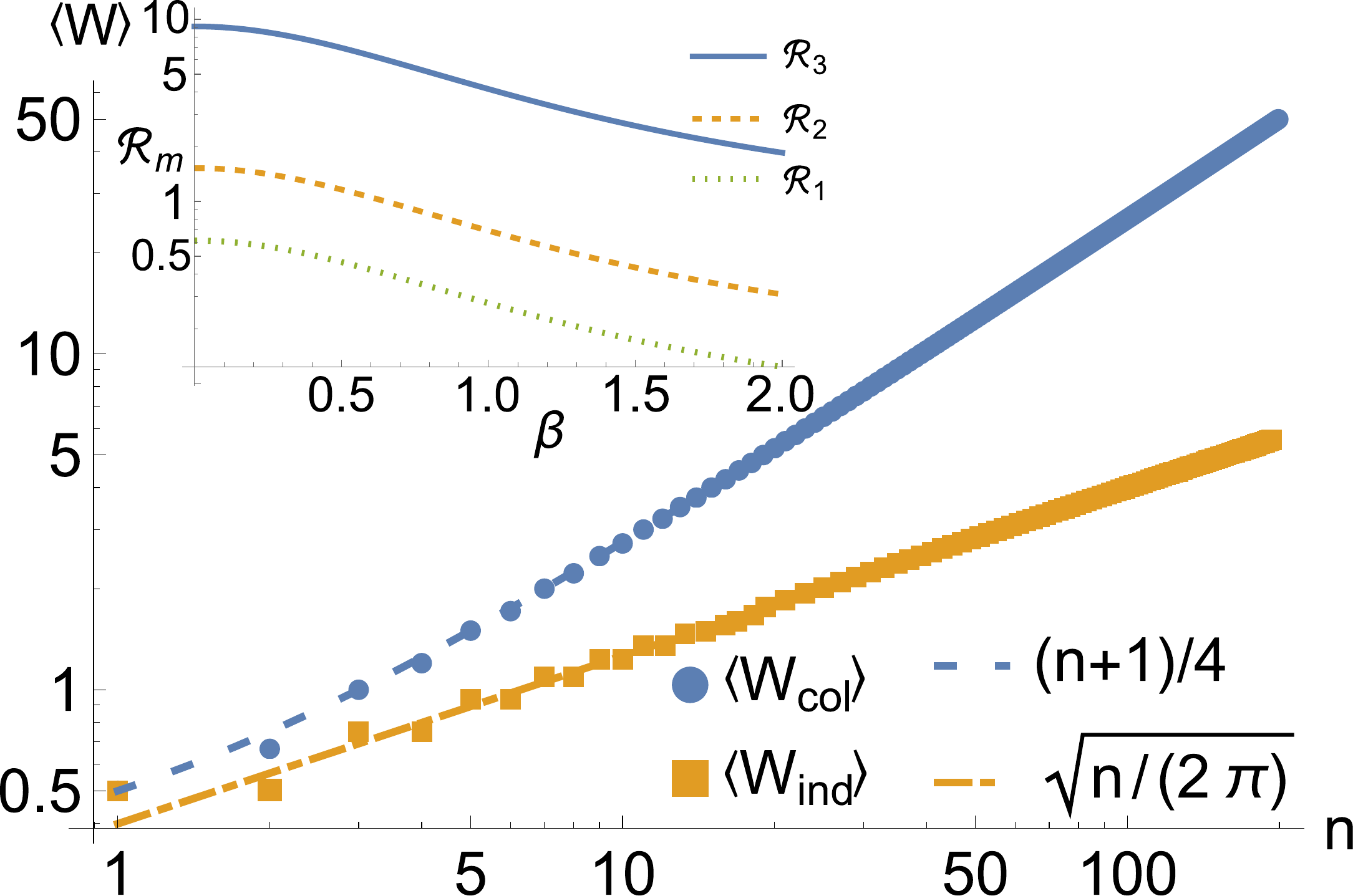} 
     \caption{The work output of the collective and independent engines 
     given in Eqs. \eqref{eq:wcol} and \eqref{eq:wind} are plotted as functions of the number of qubits $n$, for $\beta \to 0$. The dashed and dot-dashed lines represent the work output obtained in the large $n$  limit. %The inset shows the same quantities but for finite temperatures, with $\beta$ values of $0.01$ and $0.1$. 
     Inset: The figure depicts the ratios of the occupation probabilities, denoted as $\mathcal{R}_m \equiv p_m^{\rm col}/p_m^{\rm ind}$, for $m=1,2,3$ against the inverse temperature $\beta$ for a system comprising $n=6$ spin particles. The plot illustrates that the collective occupation probabilities for the higher energy levels ($m=2,3$) remain larger than the corresponding independent probabilities over a range of high temperatures. The values of $\hbar$, $k_{\rm B}$, and $\omega$ are set to 1.}
     \label{fig:w}
\end{figure}
\subsection{Information engine in the high temperature limit} \label{sec:large_temp}
In the limit $\beta=\frac{1}{k_B T}\to 0$, all the basis states will have equal probability of being occupied, and hence from Eq. \eref{eq:w}, we get the collective work output as (see Appendix),
\begin{equation} \label{eq:wcol}
\lim_{\beta \to 0}\langle W_{\rm col}\rangle =  \lim_{\beta \to 0}\sum_{m\geq0}  w_m p_m^{\rm col} \;=\; \frac{\hbar \omega}{n+1}\sum_{r=0}^{r_{\rm max}} (n-2r). 
\end{equation}
Here $p_m^{col}(\beta\to 0)=1/(n+1)$, $w_m=2\hbar\omega m = \hbar\omega (n-2r)$ with $r=n/2-m$, $r_{\rm max}=n/2$ for even $n$, and $r_{\rm max}=(n-1)/2$ for odd $n$. The above equation simplifies to $\langle W_{\text{col}}\rangle = (n+1)\hbar \omega/4$ for odd $n$, and to $\langle W_{\rm col}\rangle = \frac{n(n+2)}{4(n+1)} \hbar \omega$ for even $n$, which, in the limit of large $n$, can be approximated to $n\hbar\omega/4$. Similarly, the work output for the independent case can be obtained as
\begin{equation} \label{eq:wind}
\lim_{\beta \to 0}\braket{ W_{\rm ind} } = \lim_{\beta \to 0}\sum_{m \geq 0} p_m^{\rm ind} w_m \;=\;  \frac{\hbar \omega}{2^n}\sum_{r=0}^{r_{\rm max}} \;{ }^n C_r (n-2r),
\end{equation}
where $p_m^{ind}(\beta\to 0)=\;{ }^n C_r/2^n$ and $w_m=\hbar\omega (n-2r)$. The above equation readily simplifies to $\langle W_{\rm ind}\rangle = 2^{-(n+1)} \;{}^nC_{\frac{n+1}{2}}\; (n+1)\hbar\omega$ for odd $n$ and $\langle W_{\rm ind}\rangle = \frac{1}{2^{n+1}} \;{}^nC_{\frac{n}{2}}\; n \hbar\omega$ for even $n$.
%, where $^rC_s\equiv\frac{r!}{(r-s)! s!}$. 
By identifying the Catalan number $\mathbb{C}_r=\frac{1}{r+1} {}^{2r}C_r$~\cite{koshy2008catalan,stanley2015catalan} which approximates to $4^r/(r\sqrt{r\pi})$ in the large $r$ limit, we get $\braket{ W_{\rm ind}} \approx \sqrt{\frac{n}{2\pi}}\hbar\omega$ for both odd and even $n$. From the above, the ratio of the work output of the collective to the independent case can be calculated to be,
\begin{equation}
    \lambda_{\rm w} \;=\; \frac{\langle W_{\text{col}} \rangle }{\langle W_{\text{ind}}\rangle} \approx \frac{\sqrt{2\pi n}}{4},
\end{equation}
for large $n$ values (see Fig.~\ref{fig:wvsT}). Therefore we get a quadratic increase in the output work for the collective engine over the independent one, as also shown in Fig.~\ref{fig:w}.

We note that as discussed in ~\cite{latune20collective,jaseem2022quadratic}, in the case of a collective quantum heat engine, the improvement in its performance  is a direct consequence of the collective enhancement in the specific heat, which depends on the populations and the energies of all the magnetization $m$ states of the WM. In contrast, in the QIE introduced here, only the $m > 0$ states contribute to work extraction (see Sec. \ref{sec:infoengine}); in this case, the collective advantage stems from higher occupation probabilities $p_m^{\rm col}$ for large $m > 0$ (see the inset of Fig.~\ref{fig:w}), as compared to $p_m^{\rm ind}$ for an independent QIE, we discuss below. In the limit of high temperatures ($\beta\to 0$), the partition functions for the collective and independent cases are given by $Z_{\rm col}=n+1$ and $Z_{\rm ind}=2^n$.  With the working medium state constrained to the $j=n/2$ subspace, there is a unique eigenstate corresponding to $\ket{n/2,m}$, while for the independent case, there are ${}^nC_{n/2-m}$ degenerate eigenstates $\ket{m}$; the degeneracy  ${}^nC_{n/2-m}$ increases with decreasing $|m|$. Consequently, $p_m^{\rm ind}$ surpasses $p_m^{\rm col}$ for smaller values of $|m|$. A similar analysis for finite temperatures shows that collective advantages persist for $n \beta \ll 2/\left(\hbar \omega\right)$, as shown by the ratio $\mathcal{R}_m = p_m^{\rm col}/p_m^{\rm ind}$ for large $m > 0$ (see Eqs. \eqref{eq:pmcol}, \eqref{eq:pmind} and Fig.~\ref{fig:w}).

%The graph illustrates the probability distribution of magnetization ($m$) for both collective and independent cases for various temperatures, with $n=10$ (left) and $n=100$ (right). The range of possible $m$ values spans from $-n/2$ to $n/2$.

%%%%%%%%%%%%%%%%%%%%%%%%%%%%%%%%%%%%%%%%%%%%%%%%%%%%%%%%%%%%%%%%%%%%%%%%%%%%%
\begin{figure}[!htb]
    \centering
    \includegraphics[width=0.47\textwidth]{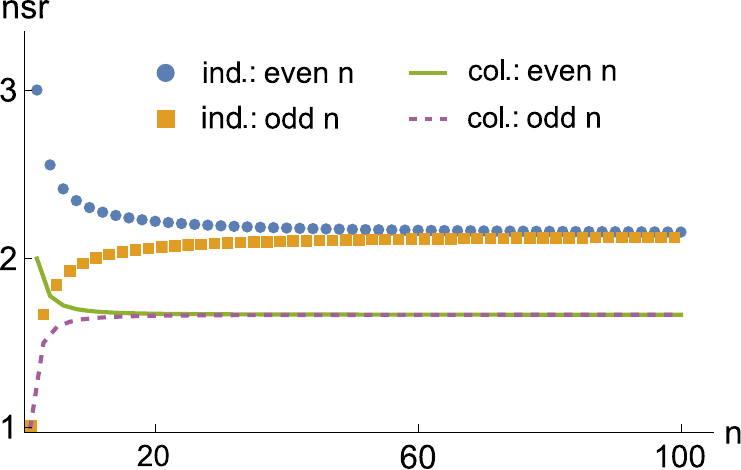} 
    \caption{The plot shows the noise-to-signal ratios for the collective (${\rm nsr}_{\rm col}$) and the independent (${\rm nsr}_{\rm ind}$) cases as functions of the number of qubits $n$, for $\beta \to 0$. %The inset presents the $\rm nsr$ as a function of $\beta$ for $n=20$ and $25$.  
    The values of $\hbar$, $k_{\rm B}$ and $\omega$ are set to 1.}
    \label{fig:nsr}
\end{figure}

%%%%%%%%%%%%%%%%%%%%%%%%%%%%%%%%%%%%%%%%%%%%%%%%%%%%%%%%%%%%%%%%%%%%%%%%%%%%%

%%%%%%%%%%%%%%%%%%%%%%%%%%%%%%%%%%%%%%%%%%%%%%%%%%%%%%%
\subsubsection{Noise to signal ratio and TUR:}
The reliability of an engine can be quantified through the noise-to-signal ratio $\text{nsr} = \text{var}(W)/\langle W \rangle^2$.
  One can calculate the variance of work for the collective and independent cases for $\beta \to 0$ as (see Eqs. \eqref{eq:varw} - \eqref{eq:pmind})
 \begin{eqnarray} \label{eq:varwcol}
\text{var}(W_{\text{col}}) &=& \frac{\hbar^2 \omega^2}{n+1}\sum_{r=0}^{r_{\rm max}} (n-2r)^2 \nonumber\\
& & \qquad - \left(\frac{\hbar \omega}{n+1}\sum_{r=0}^{r_{\rm max}} (n-2r)\right)^2,
\end{eqnarray}
and
\begin{eqnarray}
\text{var}(W_{\text{ind}}) &=& \frac{\hbar^2 \omega^2}{2^n}\sum_{r=0}^{r_{\rm max}} \;{}^n C_r (n-2r)^2 \nonumber\\
& & \quad - \left(\frac{\hbar \omega}{2^n}\sum_{r=0}^{r_{\rm max}} \;{}^n C_r (n-2r)\right)^2. 
 \label{eq:varwind}
\end{eqnarray}
As shown in Fig.~\ref{fig:nsr}, for small values of $n$, odd values yield a lower nsr compared to even values, both for collective, as well as independent engines. However, as $n$ increases, the nsrs for both even and odd values converge asymptotically, with the asymptotic value in the collective case being smaller than that in the independent case, demonstrating a collective advantage in engine performance. 

%%%%%%%%%%%%%%%%%%%%%%%%%%%%%%%%%%%%%%%%%%%%%%%%%%%%%%%%%%%%%%%%%%%%%%%%%%%%%
 \begin{figure*}[htb]
     \centering
     \subfigure[]{ \includegraphics[width=0.5\textwidth]{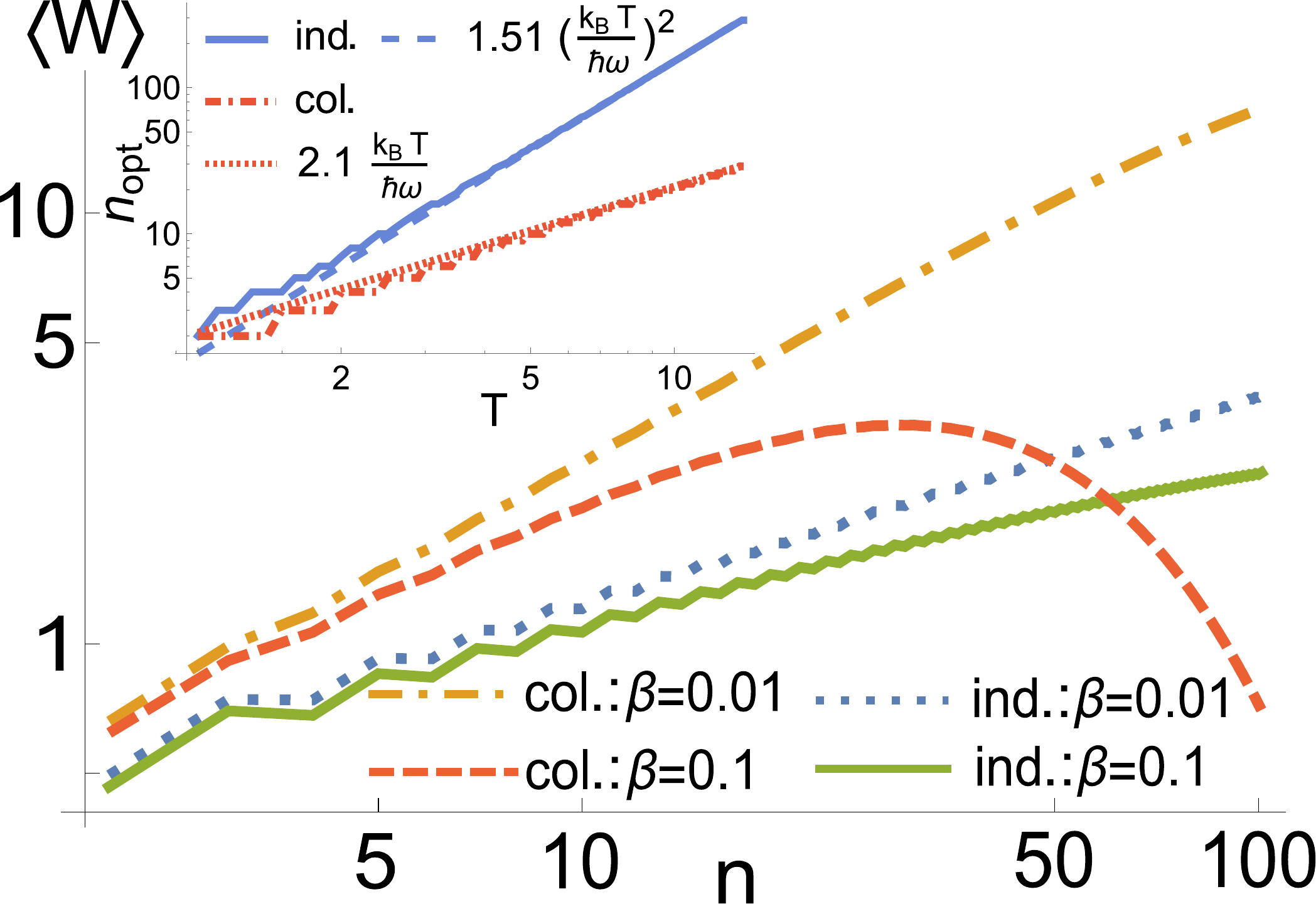} \label{fig:w_nopt}}\hfill 
     \subfigure[]{ \includegraphics[width=0.47\textwidth]{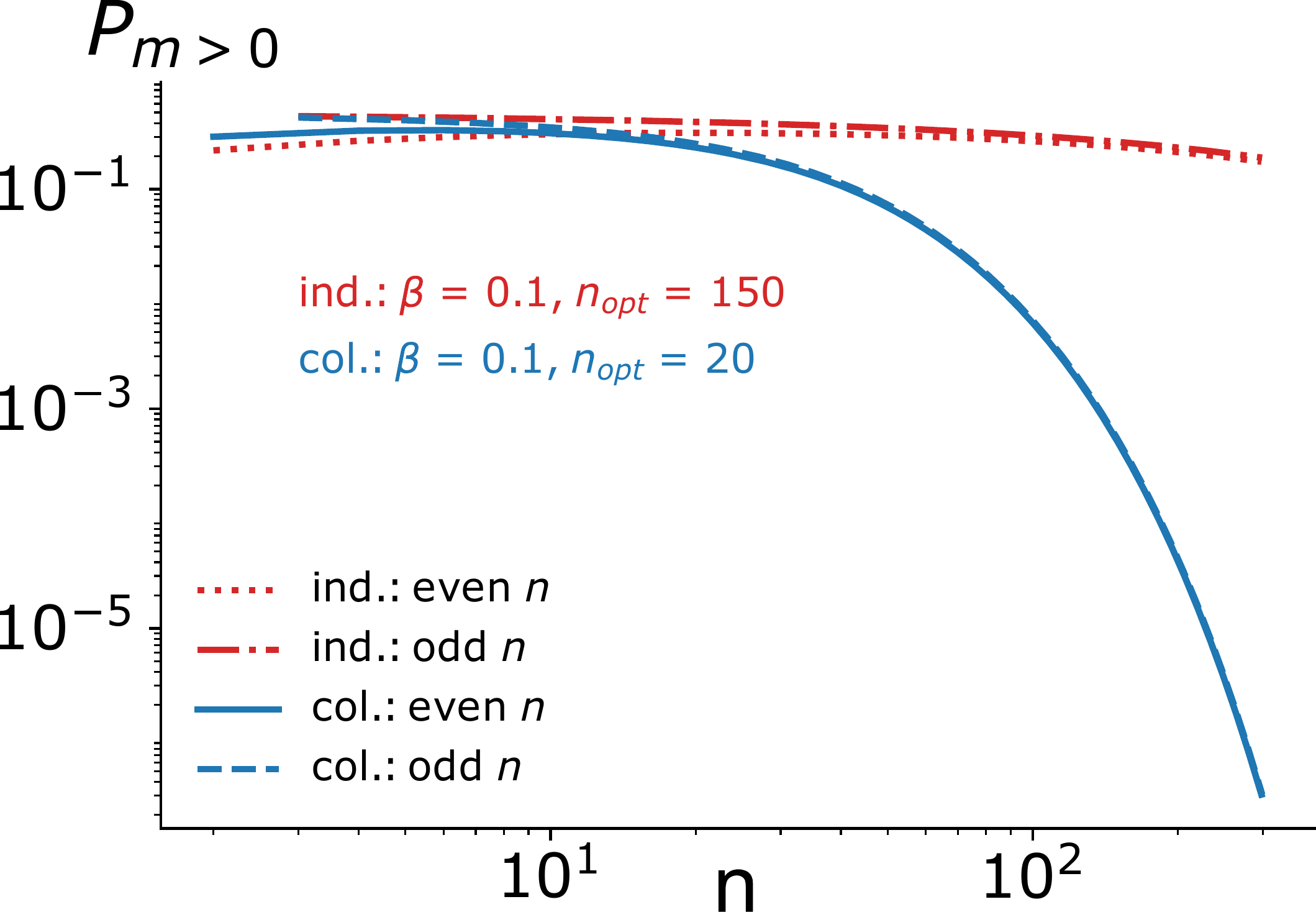} \label{figPm}}
     \caption{ (a) The work output of the collective and independent engines are plotted as functions of the number of qubits $n$, for $\beta = 0.01$ and $0.1$ (see Eqs. \eqref{eq:w}, \eqref{eq:pmcol}, and \eqref{eq:pmind}). Inset: The optimal qubit number $n_{\rm opt}^{(\alpha)}$~($\alpha = {\rm col., ind.}$), at which the work output is maximum for a given temperature, is plotted as functions of temperature $T$. Numerical fitting suggests $n_{\rm opt}^{(\rm col)} \approx 2.1 k_B T/\hbar \omega$ and $n_{\rm opt}^{(\rm ind)} \approx 1.51 \left(k_B T/\hbar \omega\right)^2$. (b) The graph depicts the probability of obtaining a positive $m$ during the measurement stroke 2, given by $P_{m>0}$, at finite temperature $\beta = 0.1$ for both collective and independent engines. The values of $\hbar$, $k_{\rm B}$, and $\omega$ are  set to 1 in both figures.} 
\end{figure*}
%%%%%%%%%%%%%%%%%%%%%%%%%%%%%%%%%%%%%%%%%%%%%%%%%%%%%%%

As discussed in \cite{Sacchi2021boson} a low nsr can be accompanied by a large entropy production $\Sigma$ in heat engines, which in turn, can be expected to reduce the efficiencies of such engines. A trade-off between nsr and $\Sigma$ can be quantified using the thermodynamic uncertainty, expressed as $\mathcal{Q} \equiv \text{nsr}\; \Sigma/k_{\rm B}$, which is generally lower bounded by the inequality $\mathcal{Q} \geq 2$ in the case of classical  \cite{barato15thermodynamic, pietzonka2018universal, horowitz2020thermodynamic} and incoherent quantum \cite{guarnieri2019thermodynamics, menczel2021thermodynamic, das2022precision} heat engines. Recently, violations of this inequality have been observed for quantum-coherent and quantum-entangled systems~\cite{agarwalla2018assessing,ptaszynski2018coherence,guarnieri2019thermodynamics,cangemi2020violation,kalaee2021violating,sacchi2021multilevel,jaseem2022quadratic}.
The behavior of $\mathcal{Q}$  in the case of information engines can be an interesting question as well \cite{potts19thermodynamic}.  In the case of the collective and independent QIEs considered here, analogous to Szilard engines \cite{kim2011quantum}, the minimum entropy production  arising due to memory erasure is given by $\Delta S_{\rm era}^{\rm min} = k_{\rm B} \ln 2$. Consequently, in contrast to what is expected for incoherent quantum heat engines \cite{guarnieri2019thermodynamics, menczel2021thermodynamic, das2022precision}, the minimum thermodynamic uncertainty $\mathcal{Q}_{\rm min} =  \ln 2\; \rm{nsr}$, which in the limit of large $n$, is less than the standard limit of $\mathcal{Q} = 2$ discussed above (see Fig.~\ref{fig:nsr}). The above result of $\mathcal{Q} < 2$ for large $n$ raises important questions regarding the validity of the standard thermodynamic uncertainty bound in the case of information engines. However, a detailed study of thermodynamic uncertainty bounds for generic QIEs is beyond the scope of the present work. Furthermore, we note that collective WM-bath coupling leads to the emergence of non-trivial steady states, signified by non-zero off-diagonal elements in $\rho^{ss}_{\rm col}$ in the local eigenbasis (see Eq. \eqref{eq:dicke_ss}) \cite{latune20collective}. This in turn have been shown to reduce the thermodynamic uncertainty $\mathcal{Q}$  below the standard bound of $2$, for quantum heat engines \cite{jaseem2022quadratic}.

\subsection{Finite temperature information engine} \label{sec_finite_temp}
%%%%%%%%%%%%%%%%%%%%%%%%%%%%%%%%%%%%%%%%%%%%%%%%
\begin{figure*}[htb]
     \centering
     \subfigure[]{\includegraphics[width=0.32\textwidth]{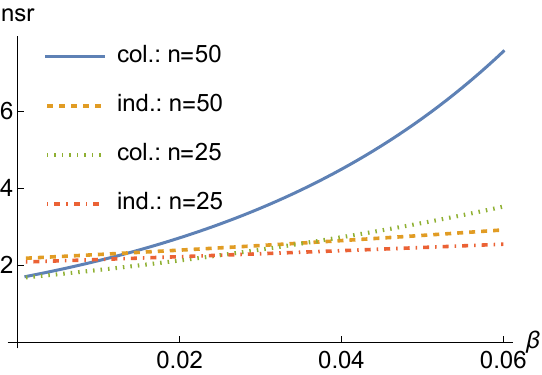} \label{fig:nsr_beta}}\hfill
     \subfigure[]{\includegraphics[width=0.33\textwidth]{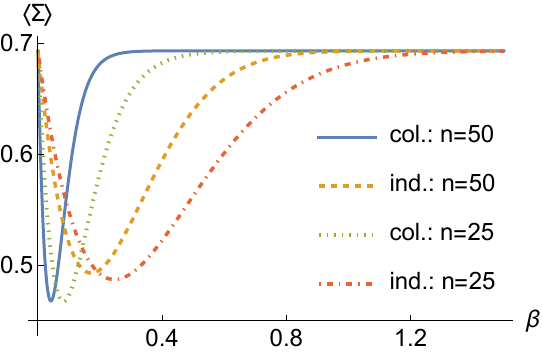} \label{fig:sigma}}\hfill
      \subfigure[]{\includegraphics[width=0.32\textwidth]{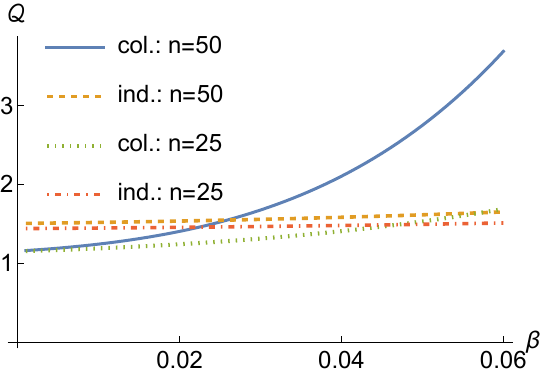} \label{fig:Q}}
    %\caption{(a) The plot presents the noise to signal ratio, $\rm nsr$ as a function of $\beta$ for $n=25$ and $50$ . (b) Plot showing the total average entropy production $\langle \Sigma \rangle$ as a function of $\beta$ for $n=25$ and $n=50$. As shown here,   $\langle \Sigma \rangle$ is always non-negative, thereby satisfying the second law of thermodynamics. (c) Plot showing the thermodynamic uncertainty $\mathcal{Q}=\text{nsr} \braket{\Sigma}/k_{\rm B}$ for $n=25$ and $n=50$. The values of $\hbar$, $k_{\rm B}$ and $\omega$ are all set to 1.}

   \caption{ The plots depict (a) the noise-to-signal ratio, $\rm nsr$, (b) the total average entropy production $\langle \Sigma \rangle$, and (c) the thermodynamic uncertainty $\mathcal{Q}=\text{nsr} \braket{\Sigma}/k_{\rm B}$ as functions of $\beta$ for $n=25$ and $n = 50$. The values of $\hbar$, $k_{\rm B}$ and $\omega$ are all set to 1.}

\end{figure*}

%%%%%%%%%%%%%%%%%%%%%%%%%%%%%%%%%%
We now focus on the finite-temperature regime. The mean and variance of the output work, as well as the nsr, can be evaluated using Eqs.~\eqref{eq:w} and \eqref{eq:varw} for arbitrary $n$ and $\beta$ (see Figs.~\ref{fig:wvsT}, \ref{fig:w_nopt}, \ref{fig:nsr_beta}). Detailed equations for calculating the mean and variance can be found in the appendix (see \eqref{app:eq_wei} - \eqref{app:eq_vco}). In the limit of large $n$, the analysis for the collective case can be simplified by replacing the summations in Eqs.~\eqref{eq:w} and \eqref{eq:varw} with integrations to get 
\begin{eqnarray}
\langle W_{\rm col} \rangle &=& \frac{[\beta  \hbar\omega  (n-2 r_{\rm max})+2]\; e^{\beta \hbar\omega r_{\rm max} } -n\beta\hbar\omega-2}{\beta  \left(e^{\beta  n \hbar\omega }-1\right)}.\nonumber\\
\label{eq:wcolfinite}
\end{eqnarray}

% %%%%%%%%%%%%%%%%%%%%%%%%%%%%%%%%%%%%%%%%%%%%%%%%%%%%%%%%%%%%%%%%%%%%%%%%%%%%%
%  \begin{figure*}[t]
%      \centering
%      \subfigure[]{ \includegraphics[width=0.47\textwidth]{nopt.pdf} \label{fig:nopt}}\hfill 
%      \subfigure[]{ \includegraphics[width=0.45\textwidth]{Pm_greater_0_beta_p1_n_300.pdf} \label{figPm}}
%      \caption{{\color{blue}(a) Optimal qubit number $n_{\rm opt}^{(\alpha)}$~($\alpha = {\rm col, ind}$), at which the work output is maximum for a given temperature, is plotted as functions of temperature $T$. Numerical fitting suggests $n_{\rm opt}^{(\rm col)} \approx 2k_B T/\hbar \omega$ and $n_{\rm opt}^{(\rm col)} \approx \pi/2 \left(k_B T/\hbar \omega\right)^2$. (b) The graph depicts the probability of obtaining a positive $m$ state, given by $P_{m>0}=\sum_{m>0}p_m$, at finite temperature $\beta = 0.1$ for both collective and independent engines. }} 
% \end{figure*}
% %%%%%%%%%%%%%%%%%%%%%%%%%%%%%%%%%%%%%%%%%%%%%%%%%%%%%%%
 
As expected, Eq.~\eqref{eq:wcolfinite} simplifies to $\langle W_{\rm col} \rangle \approx n\hbar\omega/4$ in the limit of $\beta \to 0$ (see \cref{sec:large_temp}).
Furthermore, as discussed above, the collective advantage manifests itself for high temperatures $n \beta \ll 2/\left(\hbar \omega\right)$. In contrast, numerical analysis confirms that independent engines outperform their collective counterparts for low temperatures  (see Fig.~\ref{fig:w_nopt}). 
 
We note that in accordance with the second law of thermodynamics, in an information engine operating at a temperature $T$, the maximum extractable work $\sim~k_{\rm B} T$ (see above). Again, as discussed in Sec. \ref{sec:large_temp}, in the limit of large temperatures ($k_{\rm B} T \gg n \hbar \omega$), $\langle W_{\rm col} \rangle \sim n \hbar \omega $ ($\langle W_{\rm ind} \rangle \sim \sqrt{n}  \hbar \omega$) for the collective (independent) case, which increases with increasing $n$, until $\langle W_{\rm col} \rangle \sim k_{\rm B} T$ ($\langle W_{\rm ind} \rangle \sim k_{\rm B} T$). Consequently, the optimal system size scales as $n_{\rm opt}^{\rm (col)} \sim k_{\rm B} T/ \hbar \omega$ ($n_{\rm opt}^{\rm (ind)} \sim \left(k_{\rm B} T /  \hbar \omega\right)^2$) for high temperatures for the collective (independent) case, as also verified by the inset of Fig.~\ref{fig:w_nopt}. For $n \gtrapprox n_{\rm opt}^{\rm (col)}$ ($n \gtrapprox n_{\rm opt}^{\rm (ind)}$), $\langle W_{\rm col} \rangle$  ($\langle W_{\rm ind} \rangle$) decreases with increasing $n$ (see Figs.~\ref{fig:w_nopt} and \ref{fig:wind_fall}), owing to the small probabilities $P^{\alpha}_{m>0}=\sum_{m>0}p^{\alpha}_m$  ($\alpha =$ col., ind.; see Eqs. \eqref{eq:pmcol} and \eqref{eq:pmind}) of getting positive $m$ in the measurement stroke 2 (see Fig.~\ref{figPm}).

As shown in Figs.~\ref{fig:nsr} and ~\ref{fig:nsr_beta}, collective effects result in QIEs with low nsr in the high-temperature limit, as compared to their independent counterparts. However, 
for low temperatures, independent QIEs may be more reliable than the collective ones, as signified by the nsr of collective QIEs surpassing that for independent QIEs, with increasing $\beta$, for a fixed finite $n$ (see the Fig.~\ref{fig:nsr_beta}).

% \begin{figure}[!tbh]
%     \centering
%     \includegraphics[width=0.47\textwidth]{fig5.pdf} 
%     \caption{Plot showing the total average entropy production $\langle \Sigma \rangle$ as a function of $\beta$ for $n=25$ and $n=50$. As shown here,   $\langle \Sigma \rangle$ is always non-negative, thereby satisfying the second law of thermodynamics.  The values of $\hbar$, $k_{\rm B}$ and $\omega$ are set to 1.} 
%     \label{fig:sigma}
%     \end{figure}
    
 For a finite temperature information engine, the entropy change due to the heat exchange between the WM and the heat source is given by $\Delta S^{(\alpha)}_h = - \langle W_{\alpha} \rangle/T_h$ for $\alpha = {\rm col., ind.}$, where we have considered the conservation of energy, so that on an average, the work output equals the heat input from the bath. Further, the process of memory erasure results in an increase of entropy of the universe by $\Delta S_{\rm era}\geq k_B \ln 2$. Therefore, the total entropy production is given by $\langle \Sigma_{\alpha}\rangle = \Delta S_h +\Delta S^{\alpha}_{\rm era} =-\langle W_{\alpha}\rangle/T_h+\Delta S^{\alpha}_{\rm era}$. Utilizing the results for $\langle W_{\rm col} \rangle$ and $\langle W_{\rm ind} \rangle$ (cf. Eqs. \eqref{app:eq_wei} - \eqref{app:eq_woc}), it can be shown that the entropy production is always non-negative, as depicted in Fig.~\ref{fig:sigma}, thereby satisfying the second law of thermodynamics.

Furthermore, the collective effects enhance $\langle W_{\rm col}\rangle$ at high temperatures for a given $n$, resulting in reduced entropy production ($\langle \Sigma_{\rm col}\rangle < \langle \Sigma_{\rm ind}\rangle$) and reduced thermodynamic uncertainty ($\mathcal{Q}_{\rm col} < \mathcal{Q}_{\rm ind}$) for the collective QIE, as compared to its independent counterpart. Assuming the minimal erasure entropy production, $\Delta S_{\rm era}$, to be equal to $k_{\rm B} \ln 2$, both the independent and the collective cases violate the standard TUR at high temperatures for a given $n$ (i.e., $\mathcal{Q}_{\rm col} < \mathcal{Q}_{\rm ind} < 2$), as shown in Fig.~\ref{fig:Q}. $Q_{\rm col}$ may surpass $Q_{\rm ind}$ at low temperatures, thus again showing the importance of high temperatures for achieving collective advantage in QIEs.

\section{Discussion} \label{sec_discussion}
 In this work, we have studied a many-body QIE that relies on one bit of information to extract work from a single heat bath. The engine considered here is modeled by a multi-qubit system WM interacting collectively with a heat reservoir. The collective effects observed in this system arise from the bath's inability to distinguish between the qubits as they exchange energy, which results in a non-trivial steady state of the WM. Based on the measurement outcome of the net magnetization direction of the steady state, a unitary operation may be performed to extract useful work. 

Our findings demonstrate that by utilizing effects stemming from the collective system-bath coupling, the QIE can achieve a quadratic advantage in the mean work output, as compared to its independent counterpart in the large temperature limit. 
Additionally, we have shown that this collective advantage is present in other performance metrics as well, such as low noise-to-signal ratio and low thermodynamic uncertainty, for appropriate parameter values.  However, we emphasize that these collective advantages are limited to high temperatures only ($n \beta \ll 2/\hbar \omega$). In contrast, low temperatures favor independent operation of the setup, both in terms of the mean work output (see Fig. \ref{fig:wvsT}), as well as in terms of the thermodynamic uncertainty (see Fig. \ref{fig:Q}). Our research is motivated by recent findings that demonstrate that such collective effects can significantly enhance the performance of quantum heat engines~\cite{niedenzu18cooperative, kloc19collective,  latune20collective, klimovsky19cooperative,jaseem2022quadratic}. However, in contrast to collective quantum heat engines, where the improvement in performance 
follows from collective enhancement in the specific heat of the WM, here the collective advantage stems from the higher occupation probabilities of the high positive magnetization states of the WM.

%\textcolor{blue}{Furthermore, our investigations reveal an interesting aspect in the context of finite temperatures. Specifically, when considering finite temperatures, we observe that both the work output and the thermodynamic uncertainty manifest a collective disadvantage at sufficiently large values of $n$, indicating that while collective advantages are present for various performance metrics under appropriate parameter values, there are limitations to this advantage in the context of finite temperature conditions.}

The quantum information engine proposed here can be expected to be experimentally realizable using already existing  platforms. The details of the experimental protocols will depend on the details of the particular setups used. However, we note that
collective system-bath interaction, proposed in stroke 1, has been achieved using Rydberg atoms placed in a cavity~\cite{PhysRevLett.49.117}, atomic sodium~\cite{gross1976observation}, quantum dots~\cite{scheibner2007superradiance}, and organic microcavity~\cite{quach2022superabsorption}. 
Independent system-bath coupling may be experimentally realized using nuclear spins in presence of radio-frequency fields \cite{peterson19experimental}. The measurement required in stroke 2 may be implemented by
quantum magnetometers based on nitrogen-vacancy centers in diamond, which have been shown to be suitable for performing high-precision magnetometry \cite{mamin15multipulse, balasubramanian2008nanoscale, wu2022recent}. Finally,  the unitary $U_{\rm flip}$ in stroke 3  may be realized by applying a magnetic field aligned along the $x$ axis~\cite{bodenstedt18nanoscale}. The isolation of the WM from the bath during the strokes 2 and 3 can be achieved by ensuring that both the measurement process and the unitary operation occur rapidly compared to the thermalization time of the WM~\cite{breuer2002theory}.

\section*{ACKNOWLEDGEMENTS}

V.M. acknowledges support from Science and Engineering Research Board (SERB) through MATRICS (Project No.
MTR/2021/000055) and Seed Grant from IISER Berhampur. V.M. also acknowledges Bijay Kumar Agarwalla, Gershon Kurizki and Sai Vinjanampathy for helpful discussions. 
 
%%%%%%%%%%%%%%%%%%%%%%%%%%%%%%%%%%%%%%%%%%%%%%%%%%%%%%%%%%%%%%%%%%%%%%%%%%%%%%
%\newpage
 \bibliographystyle{unsrt}
% %\color{RoyalBlue}
% \bibliography{mybib} %enters bibliography from mybib.bib

%%%%%%%%%%%%%%%%%%%%%%%%%%%%%%%%%%%%%%%%%%%%%%%%%%%%%%%%%%%%%%%%%%%%%%%%%%%%%%%%%%%%%%%%%%%%%%%%%%%
%%%      **********************      *****************************  ************************ %%%%%%
%%%%%%%%%%%%%%%%%%%%%%%%%%%%%%%%%%%%%%%%%%%%%%%%%%%%%%%%%%%%%%%%%%%%%%%%%%%%%%%%%%%%%%%%%%%%%%%%%%%

%\clearpage
\vspace{1cm}
\centerline{\bf Appendix}
\appendix*
\section{Finite temperature work and its variance}
\label{appA}
The work outputs and their variances for the independent and collective engines can be calculated from equations Eq.\eqref{eq:w} and \eqref{eq:varw}, as shown below for even and odd numbers of qubits. 
\begin{widetext}
\begin{eqnarray} 
\label{app:eq_wei}
    \braket{W_{\rm ind}^{{\rm even}\; n}} &=& n \hbar\omega  \Big[\frac{2^n \Gamma \left(\frac{n+1}{2}\right) e^{\frac{1}{2} \beta  (n+2) \hbar\omega } {\,_2F_1}\left(\frac{n}{2},n+1,\frac{n}{2}+2,-e^{\beta  \hbar\omega }\right)}{\sqrt{\pi } \left(e^{\beta \hbar \omega }+1\right) \Gamma \left(\frac{n}{2}+2\right)}-\tanh \frac{\beta \hbar \omega }{2}\Big],\\
    %2^{1-n} \hbar\omega\;  {}^{n}C_{\frac{n}{2}+1} e^{\beta  \hbar\omega }\; \text{sech} ^{n}\frac{\beta  \hbar\omega }{2}\;  \, _2F_1\left(2,1-\frac{n}{2};\frac{n}{2}+2;-e^{\beta  \hbar\omega }\right) -n \hbar\omega  \tanh \frac{\beta  \hbar\omega }{2},\\
% \end{eqnarray}
% \begin{eqnarray} 
\label{app:eq_woi}
   \braket{W_{\rm ind}^{{\rm odd}\; n}} &=&  \frac{2^n \hbar\omega  \Gamma \left(\frac{n}{2}+1\right) e^{\frac{1}{2} \beta  (n+1) \hbar\omega }  \left(2\; _2F_1\left(2,\frac{1-n}{2},\frac{n+3}{2},-e^{\beta  \hbar\omega }\right) - \;_2F_1\left(1,\frac{1-n}{2},\frac{n+3}{2},-e^{\beta  \hbar\omega }\right)\right)}{\sqrt{\pi } \Gamma \left(\frac{n+3}{2}\right) \left(e^{\beta  \hbar\omega }+1\right)^{n}} \nonumber\\
   & & -n \hbar\omega  \tanh \frac{\beta  \hbar\omega }{2},\nonumber\\
   \\
   %-n \hbar\omega  \tanh \frac{\beta  \hbar\omega }{2} +\hbar\omega  \left(e^{\beta  \hbar\omega }+1\right) \left(\sinh \frac{\beta  \hbar\omega }{2} \text{csch}\beta  \hbar\omega \right)^{n+1} \nonumber\\
  % & & \qquad \Big\{ {}^{n}C_{\frac{n+1}{2}} \, _2F_1\left(1,\frac{1-n}{2};\frac{n+3}{2};-e^{\beta  \hbar\omega }\right) +2\; {}^{n}C_{\frac{n+3}{2}} \; e^{\beta  \hbar\omega } \; \, _2F_1\left(2,\frac{3-n}{2};\frac{n+5}{2};-e^{\beta  \hbar\omega }\right) \Big\},\\
% \end{eqnarray} 
% \begin{eqnarray} 
\label{app:eq_wec}
    \braket{W_{\rm col}^{{\rm even}\; n}} &=& \frac{\hbar\omega  \left(2 e^{\frac{1}{2} \beta  (n+2) \hbar\omega }-(n+2) e^{\beta  \hbar\omega }+n\right)}{\left(e^{\beta  \hbar\omega }-1\right) \left(e^{\beta  (n+1) \hbar\omega }-1\right)},\\
\label{app:eq_woc}
   \braket{W_{\rm col}^{{\rm odd}\; n}} &=&  \frac{\hbar\omega  \left(e^{\frac{1}{2} \beta  (n+1) \hbar\omega }+e^{\frac{1}{2} \beta  (n+3) \hbar\omega }-(n+2) e^{\beta  \hbar\omega }+n\right)}{\left(e^{\beta  \hbar\omega }-1\right) \left(e^{\beta  (n+1) \hbar\omega }-1\right)},
\end{eqnarray}

\begin{eqnarray}
\label{app:eq_vie}
   {\rm var}(W)_{\rm ind}^{{\rm even}\; n} &=& \frac{1}{4} \hbar^2\omega ^2 \text{sech}^2\frac{\beta  \hbar\omega }{2} \Bigg\{ 4n+\frac{2 \Gamma (n+1) e^{\frac{\beta  n \hbar\omega }{2}}}{\Gamma \left(\frac{n}{2}+2\right) \Gamma \left(\frac{n}{2}\right)}  \;_2F_1\left(\frac{n}{2},n+1,\frac{n}{2}+2,-e^{\beta  \hbar\omega }\right)\nonumber\\
   & & \left[-\frac{2^n n \Gamma \left(\frac{n+1}{2}\right) e^{\frac{1}{2} \beta  (n+2) \hbar\omega } \;_2F_1\left(\frac{n}{2},n+1,\frac{n}{2}+2,-e^{\beta  \hbar\omega }\right)}{\sqrt{\pi } \Gamma \left(\frac{n}{2}+2\right)}+n \left(e^{\beta  \hbar\omega }-1\right)-2\right]\nonumber\\
   & & -\frac{2 n \Gamma (n+1) e^{\frac{\beta  n \hbar\omega }{2}}}{\Gamma \left(\frac{n}{2}+2\right) \Gamma \left(\frac{n}{2}\right)} \left(e^{\beta  \hbar\omega }+1\right)^{1-n} \;_2F_1\left(1,1-\frac{n}{2},\frac{n}{2}+2,-e^{\beta  \hbar\omega }\right)
   \Bigg\}
   %\hbar^2\omega ^2 e^{-(\beta  (n-2) \hbar\omega )} \left(e^{\beta  \hbar\omega }+1\right)^{-2 n} \Bigg\{\left(e^{\beta  \hbar\omega }+1\right)^n \Big[ 2 n e^{\beta  (n-1) \hbar\omega } \left(e^{\beta  \hbar\omega }+1\right)^{n-2} (n \cosh (\beta  \hbar\omega )-n+2)\nonumber\\ 
   %& & -4 \binom{n}{\frac{n}{2}+1} e^{\frac{1}{2} \beta  (3 n-2) \hbar\omega } \, _3F_2\left(2,2,1-\frac{n}{2};1,\frac{n}{2}+2;-e^{\beta  \hbar\omega }\right) \Big]\nonumber\\
  % & & -\Big[ 2 \binom{n}{\frac{n}{2}+1} e^{\beta  n \hbar\omega } \, _2F_1\left(2,1-\frac{n}{2};\frac{n}{2} + 2;-e^{\beta  \hbar\omega }\right)+n \left(e^{\frac{1}{2} \beta  (n-2) \hbar\omega }-e^{\frac{\beta  n \hbar\omega }{2}}\right) \left(e^{\beta  \hbar\omega }+1\right)^{n-1} \Big]^2\Bigg\},\nonumber\\ & & 
\end{eqnarray}

\begin{eqnarray}
\label{app:eq_vce}
   {\rm var}(W)_{\rm col}^{{\rm even}\; n} &=& \frac{\hbar^2\omega ^2 e^{\beta  \hbar\omega }}{\left(e^{\beta  \hbar\omega }-1\right)^2 \left(e^{\beta  (n+1) \hbar\omega }-1\right)^2} \Bigg\{ -n^2 e^{\beta  n \hbar\omega }+4 e^{\beta  \hbar\omega +\frac{3 \beta  n \hbar\omega }{2}}+4 e^{\frac{1}{2} \beta  (3 n+4) \hbar\omega }-(n+2)^2 e^{\beta  (n+2) \hbar\omega }\nonumber\\
   & & \qquad -4 (n+1) e^{\frac{\beta  n \hbar\omega }{2}}+4 (n+1) e^{\frac{1}{2} \beta  (n+2) \hbar\omega }+2 (n (n+2)-4) e^{\beta  (n+1) \hbar\omega }+4\Bigg\},
\end{eqnarray}

\begin{eqnarray}
\label{app:eq_vio}
     {\rm var}(W)_{\rm ind}^{{\rm odd}\; n} &=& \frac{\hbar^2\omega ^2}{\pi  \left(e^{\beta  \hbar\omega }+1\right)^2} \Bigg\{4 \pi  n e^{\beta  \hbar\omega }- \frac{\left(e^{\beta  \hbar\omega }+1\right)^{-2 n}}{2 \Gamma \left(\frac{n+3}{2}\right)^2} \Bigg\{ 2^{2 n+1} \Gamma \left(\frac{n}{2}+1\right)^2 e^{\beta  (n+1) \hbar\omega } \nonumber\\
     & & \left[n \left(e^{\beta  \hbar\omega }-1\right)\;_2F_1\left(1,\frac{1-n}{2},\frac{n+3}{2},-e^{\beta  \hbar\omega }\right)+n+1\right]^2+\pi  (n+1) \Gamma (n+1) e^{\frac{1}{2} \beta  (n+1) \hbar\omega } \left(e^{\beta  \hbar\omega }+1\right)^n \nonumber\\
     & & \left[-2 n e^{\beta  \hbar\omega } (n \cosh \beta\hbar\omega -n-2) \;_2F_1\left(1,\frac{1-n}{2},\frac{n+3}{2},-e^{\beta  \hbar\omega }\right)-(n+1)^2 \left(e^{\beta  \hbar\omega }-1\right)\right]\Bigg\}\Bigg\}
     %\frac{1}{2} n \hbar^2\omega ^2 e^{-\frac{1}{2} \beta  (n-1) \hbar\omega } \left(e^{\beta  \hbar\omega }+1\right)^{-2 (n+1)} \Bigg\{\left(e^{\beta  \hbar\omega }+1\right)^n \; \Bigg[ 8 e^{\frac{1}{2} \beta  (n+1) \hbar\omega } \left(e^{\beta  \hbar\omega }+1\right)^n \nonumber \\
    % & &  + \frac{2^n \Gamma \left(\frac{n}{2}\right) e^{\beta  n \hbar\omega }} {\sqrt{\pi } \Gamma \left(\frac{n+3}{2}\right)} {\left(n \left(n e^{2 \beta  \hbar\omega }-2 (n+2) e^{\beta  \hbar\omega }+n\right) \, _2F_1\left(1,\frac{1-n}{2};\frac{n+3}{2};-e^{\beta  \hbar\omega }\right)+(n+1)^2 \left(e^{\beta  \hbar\omega }-1\right)\right)} \Bigg] \nonumber\\
     %& & -\frac{4^n \Gamma \left(\frac{n}{2}+1\right) \Gamma \left(\frac{n}{2}\right) e^{\frac{1}{2} \beta  (3 n+1) \hbar\omega }}{\pi  \Gamma \left(\frac{n+3}{2}\right)^2} \Bigg[ n \left(e^{\beta  \hbar\omega }-1\right) \, _2F_1\left(1,\frac{1-n}{2};\frac{n+3}{2};-e^{\beta  \hbar\omega }\right)+n+1\Bigg]^2\Bigg\},\nonumber\\
\end{eqnarray}
and
\begin{eqnarray}
\label{app:eq_vco}
    {\rm var}(W)_{\rm col}^{{\rm odd}\; n} &=& \frac{\hbar^2\omega ^2 e^{\frac{\beta  \hbar\omega }{2}}}{\left(e^{\beta  \hbar\omega }-1\right)^2 \left(e^{\beta  (n+1) \hbar\omega }-1\right)^2} \Bigg\{ 4 e^{\frac{\beta  \hbar\omega }{2}}-\left(n^2+1\right) e^{\frac{\beta  \hbar\omega }{2}+\beta  n \hbar\omega }+2 (n-1) (n+3) e^{\frac{3 \beta  \hbar\omega }{2}+\beta  n \hbar\omega } \nonumber\\
    & & \qquad +e^{\beta  \hbar\omega +\frac{3 \beta  n \hbar\omega }{2}}-2 e^{\frac{1}{2} \beta  (n+2) \hbar\omega }+e^{\frac{3}{2} \beta  (n+2) \hbar\omega }+6 e^{\frac{1}{2} \beta  (3 n+4) \hbar\omega }-(2 n+1) e^{\frac{\beta  n \hbar\omega }{2}} \nonumber\\
    & & \qquad\qquad +(2 n+3) e^{\frac{1}{2} \beta  (n+4) \hbar\omega }-(n (n+4)+5) e^{\frac{5 \beta  \hbar\omega }{2}+\beta  n \hbar\omega }\Bigg\},
\end{eqnarray}
\end{widetext}
 where $\Gamma$ represents the gamma function, and $\, _pF_q(a;b;z)$ denotes the hypergeometric function, which is defined as:
\begin{eqnarray}
    {}_p F_q (a;b;z) &\equiv& {}_p F_q (a_1,...,a_p; b_1,...,b_q; z) \nonumber \\
    &=& \sum_{k=0}^{\infty}\frac{(a_1)_k...(a_p)_k}{(b_1)_k...(b_q)_k} \frac{z^k}{k!}.
\end{eqnarray}
Here the rising factorial $(x)_n$ is given by $(x)_n= x(x+1)(x+2)... (x+n-1),\; n\geq 1$. 

The above finite-temperature equations can be simplified further by assuming the large $n$-limit where the summation in Eqs.~\eqref{eq:w} and \eqref{eq:varw} can be replaced by integration, and  we get Eq.~\eqref{eq:wcolfinite} (see also Fig.~\ref{fig:w_nopt}).
As expected, Eq. \eqref{eq:wcolfinite} simplifies to $\langle W_{\rm col} \rangle \approx n\hbar\omega/4$ in the limit of $\beta \to 0$ (see \cref{sec:large_temp}).
The variance of the output work can be obtained using Eq.~\eqref{eq:varw}, where the expression for the average of the square of the work in the large $n$ limit takes the form
 \begin{eqnarray*}
    \langle W^2_{\rm col} \rangle &=& \frac{e^{\beta\hbar\omega r_{\rm max} } B - A}{\beta ^2(e^{a}-1)}, 
\end{eqnarray*}
$A = [a  (a +4)+8]$, $B= [b (b+4)+8]$, $a = \beta n \hbar\omega$, and $b= \beta  \hbar\omega  (n-2 r_{\rm max}) $.

%%%%%%%%%%%%%%%%%%%%%%%%%%%%%%%%%%%%%%%
\begin{figure}[b] 
    \centering
    \includegraphics[width=0.9\linewidth]{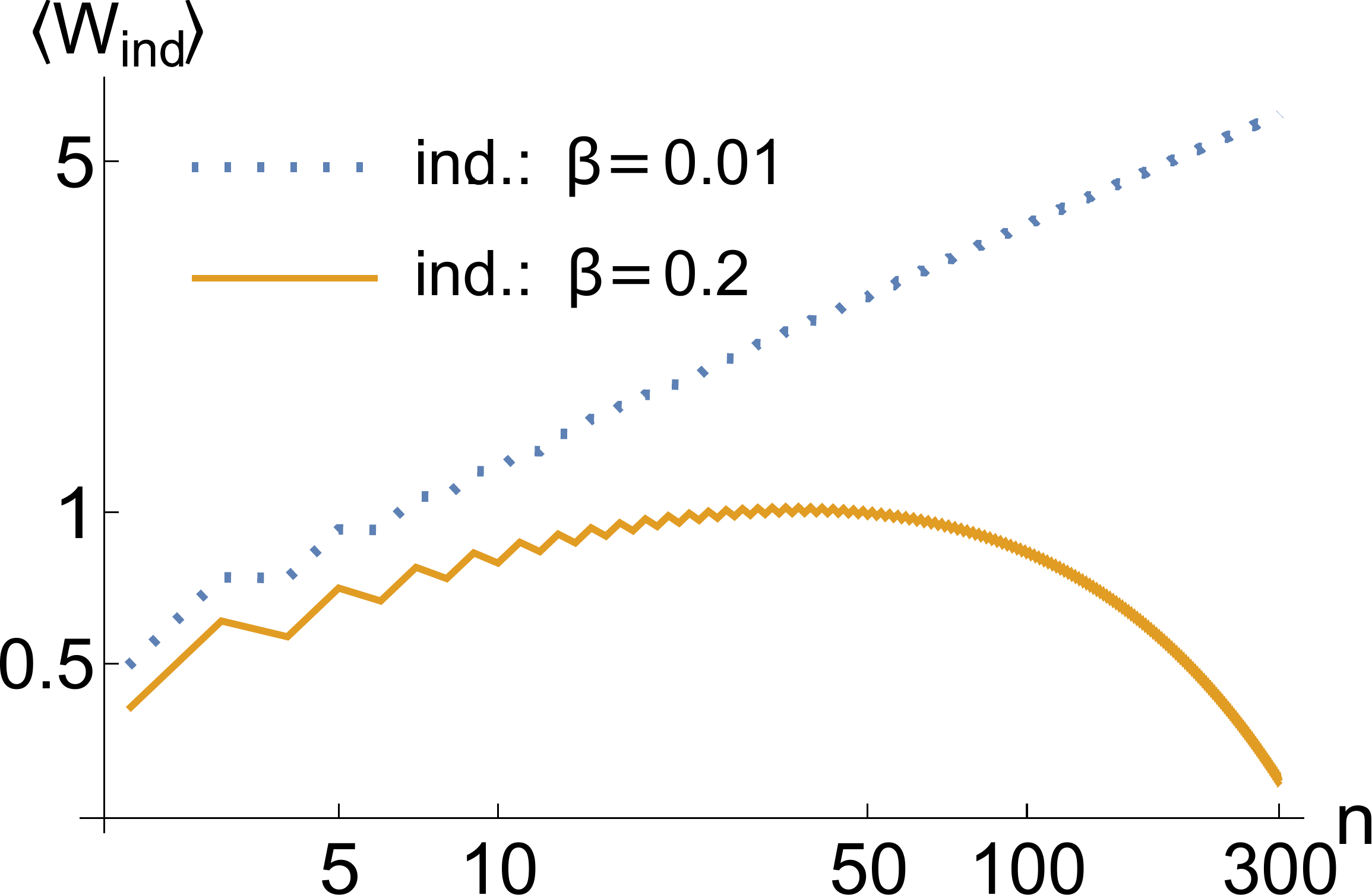}
    \caption{The graph illustrates the mean work output of the independent engine at finite temperatures, for $\beta = 0.01$  and $\beta = 0.2$. The values of $\hbar$, $k_{\rm B}$ and $\omega$ are set to 1. The plots clearly indicates a decrease in finite temperature work output for higher $n$ values, agreeing with the discussion in \cref{sec_finite_temp}.
    }
    \label{fig:wind_fall}
\end{figure}
%%%%%%%%%%%%%%%%%%%%%%%%%%%%%%%%

The mean work output $\langle W_{\rm ind} \rangle$ is shown as a function of $n$ for finite  $\beta$ in Fig. \ref{fig:wind_fall}. $\langle W_{\rm ind} \rangle$ increases with $n$ until $n_{opt} \sim \left( k_B T/\hbar \omega \right)^2$, beyond which it decreases with increasing $n$ (see Sec. \ref{sec_finite_temp} and Fig. \ref{fig:w_nopt}).

\end{document}